\documentstyle[12pt,fleqn,epsfig]{article}
\def\be{\begin{equation}}
\def\ee{\end{equation}}
\def\ba{\begin{array}}
\def\ea{\end{array}}
\def\l{\label}
\def\refe#1{(\ref{#1})}

\begin{document}

\begin{titlepage}

\rightline{\tt IC/97/75}
\rightline{\today}
\vskip2truecm

\begin{center}

{\Large\bf The $\beta$-spectrum in presence \\[2ex] 
of background potentials for neutrinos}\\[1.0cm]
{\large Francesco Vissani}
\\[0.4cm]
{\em International Centre for Theoretical Physics, 
\\ Strada Costiera 11, I-34013 Trieste, Italy}
\end{center}
\vskip0.4truecm

\noindent
\hrulefill

{\small 
\noindent 
{\sc \bf Abstract}\\
\noindent
We compute the spectrum of $\beta$-decay,  
assuming that (Majorana or Dirac) 
neutrinos propagate in constant potentials. 
We study the modifications of the spectrum due 
to the effect of these potentials.
Data on tritium decay and on  ${}^3$H--${}^3$He 
mass difference allow us to infer
bounds in the electronvolts range on the potentials.}

\noindent
\hrulefill

\vfill

\end{titlepage}
\noindent We study the modifications of the electron spectrum in 
$\beta$-decay due to hypothetical potentials 
influencing the electron neutrino propagation.
The weak interactions of the neutrinos with ordinary matter  
do modify the neutrino propagation \cite{mattereff}, but are not expected 
to be able to affect the measured $\beta$-spectrum\footnote{A density 
of scatterers $\rho$ of the order of the Avogadro number gives origin 
to a potential $V_{weak}\sim G_F \rho \sim 5\cdot 10^{-14}$ eV,
14 order of magnitudes smaller than the present experimental
sensitivity; compare however with \cite{c}.}.
In the present work we take a purely phenomenological
point of view, namely we do not attempt a study of 
the nature of the potentials, but just  
wonder which kind of observable effects they could produce. 
Apart from the interest in contemplating a theoretical possibility,
some additional motivation for this work comes from anomalies 
suggested by experimental studies of tritium decay \cite{tr,mnuexp}.

We show that potentials in the electronvolt range can lead to 
observable features in the $\beta$-spectrum,
discuss the features of the spectrum
near and far from the endpoint,
and use the experimental data to 
put an upper bound on these potentials. 

The structure of the paper is the following: 
in the first section we review the formalism for 
treating Majorana or Dirac neutrinos 
(essentially equivalent to the formalism 
put forward in \cite{p} for Dirac neutrinos);
in the second section we compute the $\beta$-spectrum 
in the presence of a background potential for the neutrino;
in the last section we discuss the features of the 
spectrum, and compare our results with the experimental findings
and with other theoretical models.

\section{Neutrino fields in presence of background potentials}
Consider a neutrino described 
by the bi-spinor $N_A$ ($A=1,2$), 
that participates in weak interactions 
and propagates in a constant potential $V_L:$
\be
{\cal L}^0_M=i \bar{N} \bar{\sigma}_a \partial^a N
-(m\, N N +{\rm h.c.}) -V_L \, \bar{N} N ;
\l{weylM}
\ee
where $a=0,1,2,3$ are Lorentz indices
and $m$ is a Majorana mass. 
The spinorial contractions\footnote{Our conventions for bispinors:
$\overline{N_A}=N_{\dot A};$ 
$N^A=\epsilon^{AB} N_B$ 
($\epsilon$ is asymmetric, and $\epsilon^{12}=1$); 
$N N=N^A N_A;$
$\bar{\sigma}_a=(1\!\!1 ,\vec{\sigma})$ ($\vec{\sigma}$ are the Pauli matrices);
$\bar{N} \bar{\sigma}_a N=\bar{N}_{\dot A} \bar{\sigma}_a^{\dot{A} B} N_B;$ 
$\bar{N} N=\bar{N}_{\dot A} N_A.$} 
imply that all terms   
but the last one are Lorentz invariants; therefore 
this latter term implies the existence 
of a preferential Lorentz frame. 
Alternatively, assuming that a second bispinor $N^c_A$ 
exists, 
we can consider the propagation described by the lagrangian: 
\be
{\cal L}^0_D=i \bar{N} \bar{\sigma}_a \partial^a N
+i \overline{N^c} \bar{\sigma}_a \partial^a N^c
-(m\, N^c N +{\rm h.c.}) -V_L \, \bar{N} N -V_R\, \overline{N^c} N^c ,
\l{weylD}
\ee
where $m$ now denotes a Dirac mass term.
$N^c$ will be assumed not to participate 
in the weak interactions. 
Eq.\ \refe{weylD} is invariant
under those U(1) transformations that act 
with opposite charges on $N$ and $N^c$
(Majorana mass terms like $N N$ and/or potential terms 
like $\overline{N^c} N$  explicitly break this invariance).  
Introducing the four-spinors: 
\be
\nu_\alpha^M=\left( \begin{array}{c}N_A\\ \bar{N}^{\dot A}\end{array}\right)
\ \ \ [{\rm Majorana}]\ \ \ \ {\rm and}\ \ \ \ \ \ 
\nu_\alpha^D=\left(\begin{array}{c} N_A\\ 
\overline{N^c}{}^{\dot A}\end{array} \right)
\ \ \ [{\rm Dirac}],
\ee
we can rewrite the lagrangians \refe{weylM} and \refe{weylD}, 
after performing  a partial integration 
in the corresponding actions, in the following form:
\be
\begin{array}{l}
\displaystyle 
{\cal L}^0_M=
\frac{1}{2}\, \overline{\nu^M}
\,[i \gamma_a \partial^a-m+\gamma^0\gamma^5 V_L]\, 
\nu^M \ \ \ \ {\rm and}\\[2ex]
\displaystyle 
{\cal L}^0_D=
\overline{\nu^D}\, [i \gamma_a \partial^a-m-\gamma^0(V_S-\gamma^5 V_D)]\, 
\nu^D  .
\l{lagras}
\end{array}
\ee
We used the gamma matrices in the chiral representation
(but of course the lagrangians \refe{lagras} have the same 
form in any representation), and set the definitions:
\be
\left\{
\begin{array}{l}
\displaystyle V_S=\frac{V_L+V_R}{2}\\[1ex]
\displaystyle  V_D=\frac{V_L-V_R}{2}  .
\end{array}
\right. 
\l{vdef}
\ee

The  solutions of the equations of motion are easily obtained.
In the Majorana case we search for the 
positive energy solutions of the form:
\be
\nu^M(x)=\exp[-i E(\lambda) t+i\vec{p}\cdot \vec{x}]\times u(\lambda,\vec{p}) ;
\l{ans}
\ee
where $\lambda=\pm 1$ denotes the helicity: 
$\vec{\Sigma}\cdot \vec{p}\,/\, p\, [ u(\lambda,\vec{p}) ]=\lambda\, u(\lambda,\vec{p}),$
using the notation $p$ for the squared three-momentum. 
Since the spinor $u$ obeys:
\be
[E(\lambda) \gamma^0 - m 
- \gamma^0\gamma^5 (\lambda p-V_L) ]\, u(\lambda,\vec{p})=0
\l{eqom}
\ee
the potential can be taken into account replacing 
\be
p \to p-\lambda V_L
\l{prepl}
\ee
in the usual formulae for energy and for 
spinors.
In particular, the dispersion relation, 
depicted in fig.\ 1, becomes:
\be
E(\lambda)=\sqrt{m^2+(\lambda p-V_L)^2} .
\l{disp}
\ee
The steps (\ref{ans}-\ref{disp}) can be
repeated for the Dirac field, since 
after the simple replacement:
\be
V_L\to V_D 
\label{vrepl}
\ee 
the wave $\nu'(x)=\exp(iV_S t)\; \nu^D(x)$ is shown to obey 
the same equation of $\nu^M(x)$.

The quantization of the Majorana field yields us:
\be
\nu^M(x)=\sum_{\lambda} \int \frac{d\vec{p}}{2 E(\lambda) (2\pi)^3}
\begin{array}[t]{l} \left[ a(\lambda,\vec{p})\ u (\lambda,\vec{p})
\ {\rm e}^{-i E(\lambda) t 
+i \vec{p}\cdot \vec{x}} \right. \\ 
\left. + \xi
a^\dagger(\lambda,\vec{p})\ C {\overline{u(\lambda,\vec{p})}\,}^t\ 
{\rm e}^{+i E(\lambda) t -i \vec{p}\cdot \vec{x}} \right] ,
\end{array}
\label{quantM}
\ee 
where spinors and operators are normalized to $2E(\lambda),$ and
$\xi$ is a phase factor entering the Majorana condition:
${C\overline{\nu^M(x)}\, }^t=\xi \nu^M(x).$ 
In order to obtain the quantized Dirac neutrino $\nu^D$ 
one has $(i)$ to replace the creation operator $a^\dagger$ in \refe{quantM}
with a different  degree of freedom $b^\dagger,$ $(ii)$ perform 
the substitution \refe{vrepl} into the expressions for energy
and for spinors, and finally $(iii)$ 
multiply the resulting field $\nu'$ by the phase $\exp(-iV_S t).$   
Due to this last step, $V_S$ will enter in  
the expression for the energy of a Dirac 
neutrino (resp.\ antineutrino) as: $E(\lambda)+V_S$  
(resp.\ $E(\lambda)-V_S$).  

\section{$\beta$-spectrum}
 
The $\beta$-decay is due to 
the charged current lagrangian:
\be
{\cal L}_{CC}=-
\frac{G_F\cos\theta_C}{\sqrt{2}} \ 
\bar{u}\gamma^a(1-\gamma^5)d\ 
\bar{e}\gamma_a(1-\gamma^5)\nu ,
\l{fermi}
\ee
where $\nu=\nu^M$ or $\nu^D.$ 
The potentials modify the $\beta$-spectrum. 
In this section we describe in detail the modifications which 
take place for a Majorana neutrino, 
and indicate how to adapt the formulae to the Dirac case.
To be concrete, we will focus on the 
tritium decay: 
\be 
{}^3{\rm H} \to {}^3{\rm He}+e+\overline{\nu},\ \ \ \ \ \ \ \ \ \ 
Q=18590\ {\rm eV}\  ({\rm ref.}\ \cite{deltam})
\ee
for which the energy release $Q$ 
(the nuclear mass difference being $\Delta{M}=m_e+Q$)
is so small that all particles 
except possibly the antineutrino can be
considered non-relativistic.

The matrix element squared is:
\be
|{\cal M}(\lambda)|^2={\rm const.}\times 
(E_\nu(\lambda) + \lambda p_\nu - V_L) ,
\label{matrix}
\ee
where the constant depends on the fundamental parameters in \refe{fermi},
on the nuclear matrix element and on the masses of the electron
and of the nuclei
(one simple derivation of this result: 
consider the usual result for $|{\cal M}|^2$ for neutron decay,  
substitute $(p_\nu)^a$ with $(p_\nu)^a+\lambda m_\nu\, (n_\nu)^a$
to account for antineutrino 
polarization, perform the replacement \refe{prepl}
and then take the non-relativistic limit). 
For the Dirac case, use \refe{vrepl}.
As expected, the matrix element corresponding to the 
positive helicity dominates in the relativistic limit.

Since the antineutrino energy depends on the
polarization, the maximal electron energy
will also depend on it. 
The minimal antineutrino energy is  
$E^{\rm \scriptscriptstyle min}_\nu=
m_\nu$ if $\lambda={\rm sign}(V_L),$  or 
$E^{\rm \scriptscriptstyle min}_\nu=
\sqrt{m_\nu^2+V_L^2}$ if $\lambda=-{\rm sign}(V_L).$
Therefore, from the energy conservation condition  we obtain:
\be
E_e^{\rm \scriptscriptstyle max}(\lambda)= \left\{ 
\begin{array} {lll}
{\Delta M}- m_\nu & &{\rm if}\ \lambda={\rm sign}(V_L)\\
{\Delta M}- \sqrt{m_\nu^2+V_L^2} &\ \ \ \ \ \ \ \ \  &{\rm if}\ 
\lambda=-{\rm sign}(V_L) .
\end{array}
\right.
\label{endpoint}
\ee 
The first case corresponds to
a non-zero minimum momentum of the antineutrino:
$p_\nu=|V_L|.$
For the Dirac neutrino, in order to obtain the maximum electron energy
one has $(i)$ to perform the replacement \refe{vrepl}
and $(ii)$ to add $V_S$ in the previous formula.

Let us consider the phase space of the reaction.
Using the delta function from momentum conservation
to integrate away the helium-3 three-momentum, 
integrating over the possible direction of the electron, and
averaging over those of the antineutrino we get:
\be
d\Phi_3(\lambda)= 
\frac{1}{(2\pi)^3} \frac{p_e^2 d p_e}{ 2 E_e E_{\,{}^3 {\rm He}}
}\ F_\nu(\lambda),
\ \ \ \ \ \ \ 
F_\nu(\lambda)=
\frac{p_\nu^2 d p_\nu}{ E_\nu(\lambda)} \delta({\Delta M}-E_e-E_\nu(\lambda)) 
\label{phase}
\ee
where $F_\nu(\lambda)$ is a phase space factor 
related to the antineutrino degrees of freedom.  
When integrating over the unobservable antineutrino momentum,
one has to take into account the dependence of the energy
on the helicity:

\noindent (1) If $E_e<{\Delta M}-\sqrt{m_\nu^2+V_L^2}$ 
(corresponding to $E_\nu> \sqrt{m_\nu^2+V_L^2},$ see fig.\ 1), 
there are two possible momenta, one for each helicity state:
$p_\nu=\sqrt{E_\nu^2-m_\nu^2}+\lambda V_L.$
For them we get: 
\be
F_\nu(\lambda)=
\left. \frac{ (\sqrt{E_\nu^2-m_\nu^2} +
\lambda V_L)^2}{\sqrt{E_\nu^2-m_\nu^2}}
\right|_{E_\nu={\Delta M}-E_e} .
\label{Flow}
\ee
(2) In the case $E_e>{\Delta M}-\sqrt{m_\nu^2+V_L^2}$ there is 
only one helicity state, $\lambda={\rm sign}(V_L),$ but
there is a second value of the momentum at fixed
energy:
$p_\nu=$ $-\sqrt{E_\nu^2-m_\nu^2}$ $+\lambda V_L.$ The phase 
space factor for this determination of the momentum is:
\be
F_\nu(\lambda)=
\left. \frac{ (\sqrt{E_\nu^2-m_\nu^2} -
\lambda V_L)^2}{\sqrt{E_\nu^2-m_\nu^2}}
\right|_{E_\nu={\Delta M}-E_e} ,
\label{Fhigh}
\ee
whereas, for the other momentum, formula \refe{Flow} 
applies (with the same value of the helicity).
Notice that this phase space factor is 
formally identical to the one that we would obtain 
using the $\lambda=-{\rm sign}(V_L)$ and 
$p_\nu=\sqrt{E_\nu^2-m_\nu^2}+\lambda V_L,$
even if, strictly speaking, this ``square momentum'' $p_\nu$
is negative.

The differential spectrum of the electron can be estimated by
summing up the antineutrino helicities,
$d\Gamma=\sum_\lambda d\Phi_3(\lambda)$ 
$|{\cal M}(\lambda)|^2\,/\, (2\ m_{\, {}^3{\rm He} }),$
where as discussed above, both 
the matrix elements \refe{matrix}
and the phase space \refe{phase}
depend on the helicity: 
\be
\frac{d\Gamma}{d E_e}\propto
p_e E_e  F(Z,E_e) \times \!\!
\left. \sum_{\sigma=\pm 1}\! 
\frac{E_\nu+\sigma\, {\cal P}_\nu}{2\ {\cal P}_\nu}
\ ({\cal P}_\nu +\sigma\, V_L)^2 \, \right|_{E_\nu=\Delta M -E_e} ,
\label{diffM}
\ee
in this formula $p_e=\sqrt{E_e^2-m_e^2}$ and ${\cal P}_\nu=\sqrt{E_\nu^2-m_\nu^2};$ 
$F(Z,E_e)$ embodies the electromagnetic corrections
due to the interaction of the final states.
The maximal electron energy is 
\be
E_e^{\rm \scriptscriptstyle endp.}={\Delta M}-m_\nu \ \ \ \ [{\rm Majorana}],
\l{endM}
\ee
and is not shifted by the potential $V_L$
(we recall that $\Delta M$ denotes the
${}^3$H--${}^3$He mass difference).

In the Dirac case we get:
\be
\frac{d\Gamma}{d E_e}\propto
p_e E_e  F(Z,E_e) \times \!\!
\left. \sum_{\sigma=\pm 1}\! \frac{E_\nu+\sigma\, 
{\cal P}_\nu}{2\ {\cal P}_\nu}
\ ({\cal P}_\nu +\sigma\, V_D)^2 \, \right|_{E_\nu=\Delta M -E_e-V_S} .
\label{diffD}
\ee
The maximal electron energy is 
\be
E_e^{\rm \scriptscriptstyle endp.}=\Delta{M}-m_\nu+V_S\ \ \ \ [{\rm Dirac}],
\l{endD}
\ee
and therefore differs from the Majorana case if $V_S\neq 0.$

The Kurie plot is defined as:
\be
K(E_e)= \sqrt{\frac{1}{p_e E_e F}\,\frac{d\Gamma}{dE_e}};
\ee
if masses and potentials are absent, 
$K(E)$ turns out to be a line approaching 
zero at the endpoint.  
In the case of small neutrino mass  
the potential enters the formula for the Kurie plot as:
\be
K(E_e)\propto |\Delta M -E_e+V_L|,\ \ \ \ \ {\rm small}\ m_\nu
\l{kur}
\ee
for both Majorana and Dirac cases, but the 
endpoint is in a different position unless $V_S=0$
(compare eqs.\ \refe{endM} and \refe{endD}).

The integral spectrum, often used in the
experimental analysis is: 
\be
I(E_e)=\int^{E_e^{\rm \scriptscriptstyle endp.}}_{E_e}
\frac{d\Gamma}{dE_e}\ dE_e .
\label{int}
\ee

\section{Discussion of the results}

We first discuss some typical features of the
$\beta$-spectrum, and then compare them with the experimental
results and with other theoretical models.

\subsection{Features of the $\beta$-spectrum}

We start with some general remarks 
on the differential spectrum, 
eqs.\ \refe{diffM} and \refe{diffD}:
\begin{enumerate}
\item While the effects of the masses
on the spectrum are of order $(m_\nu/E_\nu)^2,$  
the effects of the potential are of order $V/E_\nu.$ 
\item The Majorana and the Dirac neutrino spectra   
for equal masses are identical,  not only if 
the potentials are absent, but under the weaker assumption  
$V_S=0$ (which, by definition \refe{vdef}, is equivalent  to $V_D=V_L$).
\item There is a curious behaviour of the differential spectrum 
at the endpoint:  If the mass and the potential are 
both non-zero, it goes to infinity,
since ${\cal P}_\nu\to 0$ in eqs.\ \refe{diffM} and \refe{diffD}.
This fact however does not imply contradictions with the physical
interpretation, since in any given finite interval the 
probability of transition is finite.
This  behaviour is due to the increasing number of states 
corresponding to a fixed energy interval when 
the neutrino energy decreases toward its minimum, compare
with eqs.\ (\ref{phase}-\ref{Fhigh}), or see fig.\ 1
(we recall that the minimum corresponds 
to the--infinite--set of states $\vec{p}_\nu=|V_L|\, \vec{n}_\nu,$ 
where the orientation $\vec{n}_\nu$ is arbitrary)\footnote{If 
the potential $V_L$ is zero, the minimum
shifts at $p_\nu=0,$ and therefore the phase space factor
$d\vec{p}_\nu$ shrinks to zero when the minimum is approached, 
whereas if $m_\nu=0$ 
the dispersion relation curve in this limit does not flatten,
so that the phase space attains a finite value.}.
\end{enumerate}
Continuing the discussion of the differential spectrum,  
we come to the differences between the Majorana and the Dirac cases.
For a Majorana neutrino it is sufficient to
distinguish if $V_L$ is larger or smaller than zero.  
For small mass, in the first case 
the spectrum just translates upward, whereas in 
the second case it translates downward for energies
$E_e<\Delta M+V_L,$ where it becomes zero, and then
it increases up to the endpoint \refe{endM}: a dip is present.
These features are clearly illustrated by eq.\ \refe{kur}.
The Dirac case is identical, apart for the fact 
that the position of the endpoint depends on $V_R.$ 
Notice in particular that for $V_L=V_R$ ($=V_S,$ $V_D=0$ 
due to eq.\ \refe{vdef}) the Kurie plot
is simply a straight line which approaches 
zero at $E_e=\Delta M +V_L.$

Let us conclude with some comments 
on the integral spectrum. 
In the Majorana case, 
the integral spectrum  for $V_L>0$ moves to the
left, and approaches in a steeper fashion the endpoint 
when the mass is increased. Instead if $V_L<0,$ 
for masses of the order of $|V_L|$ or smaller 
the integrated spectrum remains approximatively
constant in the region that extends 
$2\, |V_L| $ below the endpoint (the size of the dip
in the differential spectrum). 
In the Dirac case for positive and increasing 
values of $V_R$ the endpoint moves to the right,
extending the region in which the
integral spectrum stays approximatively constant,
in accord with the features already discussed.

\subsection{Comparison with the experimental results}

Experimentally one can study the shape 
of the $\beta$-spectrum in the region
far from the endpoint, or near the endpoint. 
Let us discuss these two regions in turn.  

\subsubsection{Region far from the endpoint}
Due to the first feature of the spectrum 
emphasized in the previous subsection,  
a mass of some eV would 
give a deviation of the spectrum only near the endpoint, 
whereas a potential $V_L$ of few eV 
would give a detectable 
shift of the spectrum also in the region
far from the endpoint (the eV range being the 
present experimental accuracy).
In this hypothesis, the 
experiments which measure
the ${}^3$H--${}^3$He mass difference 
fitting the $\beta$-spectrum far from the endpoint 
are sensitive to $\left. \Delta M \right|_\beta=\Delta M +V_L$
(eq.\ \refe{kur} applies since the 
neutrino energy is much larger that its mass in the
region under discussion) 
which would systematically deviate  
from the $\left. \Delta M \right|_{{\rm non}-\beta},$ 
obtained by techniques of measurement of the mass difference 
which do not rely on the $\beta$-spectrum.
Whereas older data were to some extent indicative of 
a discrepancy between the $\beta$- and the 
non-$\beta$-determinations of $\Delta M,$ 
more recent and precise determinations are in remarkable  
agreement \cite{deltam2}.
Including the six more recent 
measurements reported in ref.\ \cite{deltam2}
(5 $\beta$- and 1 non-$\beta$-techniques) 
we estimate: 
\be
V_L = \left. \Delta M \right|_{\beta}
-\left. \Delta M \right|_{{\rm non}-\beta}=0.3 \pm 2.7 \ {\rm eV}
\l{vlbound}
\ee
Eq.\ \refe{vlbound} rules out  $V_L>5.6$ eV 
or $V_L<-5.0$ eV at 95\% CL. 

\subsubsection{Endpoint region}
The region near the endpoint is very interesting for 
the search of electron antineutrino mass. 
Fits to recent data \cite{tr,mnuexp} suggest a 
``negative masses squared'', $m_\nu^2\approx -20$ eV$^2$.
Recalling the well-known feature 
that a (positive) neutrino mass term 
depletes the endpoint region of the differential spectrum,  
one realizes that a ``negative mass squared'' 
simply amounts to the fact that the measured spectrum 
is even larger than what one 
would expect postulating a massless neutrino. 
Putting aside the possibility of statistical or 
instrumental effects,
such findings could be taken as suggestions of 
a peak-like structure near the endpoint region.

In the following we focus on the possibility to use   
the model under discussion to explain this feature. 
But before doing so, let us spend a few words of 
warning:
\begin{itemize}
\item Even if certain choices of 
the potentials are able to 
explain the presence of a peak in the 
endpoint spectrum, one should
keep in mind that different types of patterns can be obtained 
with other choices of potentials (and, in particular,
of their signs). In addition, other models exist which  
may entail modifications of the endpoint spectrum 
(see also the discussion below).
\item The fit value: $m_\nu^2\approx-20$ eV${}^2$ 
quoted above is dominated by 
the experimental findings of the 
Troitsk collaboration \cite{tr}, which 
reports the smallest errors, 4.8 and 5.8 eV$^2.$ 
The confidence level of the average fit values 
of \cite{tr} and \cite{mnuexp} 
is of 0.6\% or 20\%, if we sum quadratically or linearly
the statistical and the systematic errors
(when they are quoted separately). 
These considerations suggest that it will be quite 
important to follow 
future experimental developments. 
\end{itemize}

As it is clear from eq.\ \refe{kur}, a
Majorana neutrino with a small mass
and a {\em negative} potential $V_L$ of few eV could 
explain the appearance of a peak
in the differential spectrum exactly at the endpoint,
and a dip $|V_L|$ electronvolts below. 
But for a Majorana neutrino 
the position of the dip is linked to a downward 
shift of the spectrum in the region far from the endpoint, 
and therefore  the bound \refe{vlbound} applies.

This limitation does not hold necessarily for a Dirac neutrino.
In fact, if $V_L\sim 0$  and $V_R$ is {\em positive},
there is no deviation in the differential spectrum
up to the point in which it vanishes,
$E_e=\Delta M,$ but for larger energies the spectrum would start to 
rise constantly, up to the endpoint $E_e=\Delta M+V_R/2$ 
(we assumed again a small neutrino mass term).
We conclude that a potential $V_R$ of some eV, 
felt by a Dirac neutrino, can manifest itself in a peak of the 
differential spectrum near the endpoint region.
A $V_R$ in the twenty eV range or so 
can already be excluded by the data. 
Let us stress that in such a scheme there is no need 
to attach any physical meaning 
to the ``negative mass squared''. 

\subsection{Comparison with other theoretical models}
An observable deviation of the spectrum in the region  
far from the endpoint region  (eq.\ \refe{vlbound} 
and related discussion), 
characterizes the scenario  in which the neutrino propagates in
a (large) background potential $V_L.$ 

Instead, peak-like structures in the endpoint region 
can also arise in the presence of a degenerate 
neutrino sea, filled up to the Fermi level $\mu_F$ \cite{w}. 
In fact, the reaction of absorption of a neutrino 
$\nu_{\rm sea} + {}^3{\rm H}\to e + {}^3{\rm He}$ 
permits the emission 
of electrons with energies between $\Delta M$ and $\Delta M+\mu_F,$
if the neutrino mass is supposed to be 
small in comparison with $\mu_F.$
Therefore, one cannot discriminate between 
these two theoretical possibilities from the experimental  
information on the $\beta$-spectrum. 
Instead, if neutrino masses are not small 
it is possible, at least in principle, 
to distinguish between the two cases 
({\em i.e.}\ potential {\em versus} chemical potential).
In fact, if we suppose the existence of 
a degenerate sea of massive neutrinos,   
no electron could be emitted in the energy range 
between $\Delta M-m_\nu$ and $\Delta M+m_\nu$ 
(the emission due to the neutrino absorption  
taking place for larger energies, up to
$E_e=\Delta M+m_\nu+\mu_F$) \cite{ih}.
Such a discontinuity of the differential 
spectrum is absent in the models 
discussed in the present work.

\vskip0.5cm
\noindent{\Large\bf Acknowledgments}
\vskip0.3cm
\noindent
I wish to thank A.Yu.\ Smirnov who suggested 
the question studied in the present work, 
and to acknowledge discussions with
S.\ Cherubini, S.\ Esposito and S.\ Pakvasa.

\newpage

\begin{figure}
\centerline{\epsfig{file=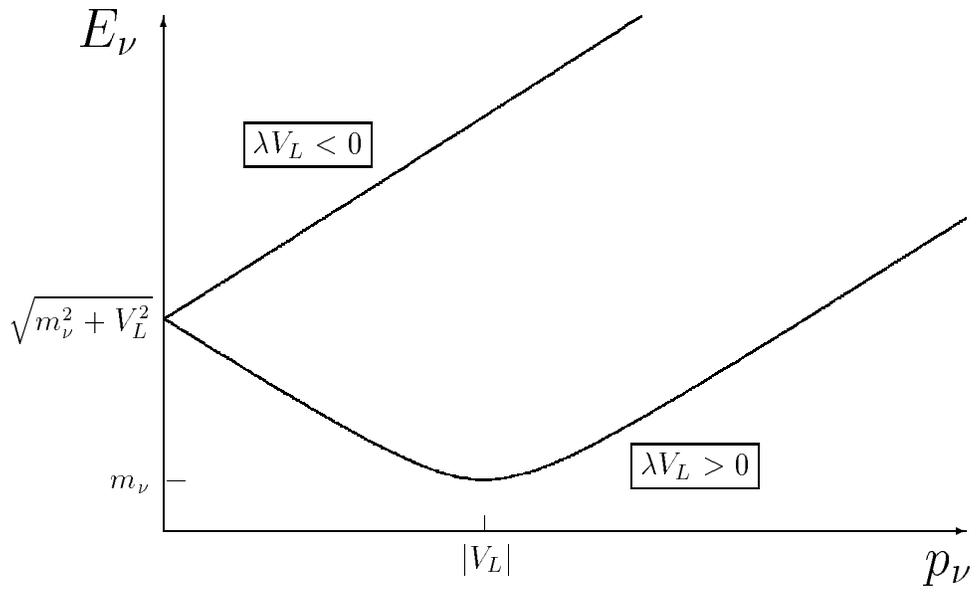}}
\vskip1truecm
\caption{Dispersion relation for Majorana neutrinos 
in presence of a constant potential $V_L.$}
\end{figure}


\begin{thebibliography}{99}
\bibitem{deltam2}
See for example table I$\!$I 
of the previous reference,
in which the historical excursus 
of the determinations of $\Delta M,$ 
and the references to the previous works
are presented.
\end{thebibliography}
\end{document}